\documentclass[11pt, epsfig]{article}
\usepackage{epsfig, amsmath, amssymb, amsthm}
\def\Na{{\rm I}\!{\rm N}}
\def\lbl{\label}
\def\be{\begin{equation}}
\def\ee{\end{equation}}
\def\lbl{\label}

\title{On transition to bursting via deterministic chaos}
\author{ Georgi S. Medvedev
\thanks{
Department of Mathematics, Drexel University, 3141 Chestnut Street,
Philadelphia, PA 19104, {\tt medvedev@drexel.edu} }
}

\begin{document}
\maketitle

\begin{abstract}
We study statistical properties of the irregular bursting arising in 
a class of neuronal models close to the transition from
spiking to bursting. Prior to the transition
to bursting, the systems in this class develop chaotic attractors, which
generate irregular spiking. 
The chaotic spiking gives rise to irregular bursting. The duration
of bursts near the transition can be very long. We describe the statistics 
of the number of spikes and the interspike interval distributions  within one burst 
as functions of the distance from criticality.
\end{abstract}

\begin{figure}
\begin{center}
\epsfig{figure=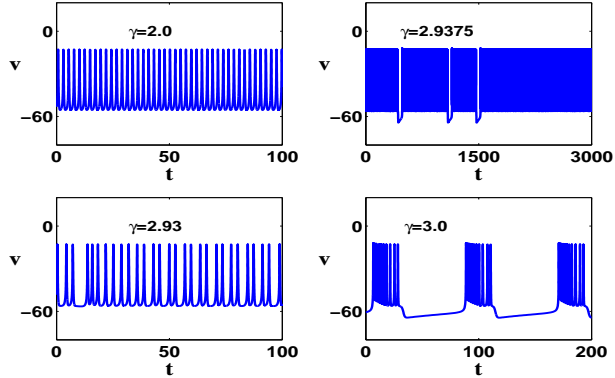, height=2.0in, width=3.2in, angle=0}
\end{center}
\caption{
Periodic and aperiodic firing patterns generated by (\ref{1})-(\ref{3}) 
in the regimes close to the transition from spiking to bursting.
The values of parameters
are $C=1$ $\left(\mu F cm^{-2}\right);$ $g_{NA}=20$, $g_K=10$, 
$g_L=8$ $\left(mS cm^{-2}\right)$; $E_{Na}=60$, $E_K=-90$, 
$E_L=-80$ $\left(mV\right)$; 
$a_m=-20$, $a_n=-25$, $a_w=-20$ $\left(mV\right)$;
$b_m=15$, $b_n=5$, $b_w=5$; $\tau_n=0.152$, $\tau_w=20$ $\left(ms^{-1}\right)$,
and $I=5 pA$. }
\label{f.1}
\end{figure}

\setcounter{equation}{0}
Bursting oscillations are ubiquitous in the experimental and modeling studies
of excitable cell membranes. Many models generating bursting have been
subject to intensive research due to their physiological 
significance and dynamical complexity (see \cite{chay84, IZH05, M05, rinzel87, 
RE89, SC04, terman92} and references therein).  
Under the variation of parameters even minimal $3D$ models of bursting
neurons exhibit a rich variety of periodic and aperiodic dynamical patterns
corresponding to different spiking and bursting regimes. 
The transitions between these patterns may contain complex dynamical
structures such as period-doubling cascades and deterministic chaos.
In particular, it was shown that the transition from tonic spiking to bursting
in a class of bursting neuron models, so-called square-wave bursters, 
contains windows of chaotic dynamics \cite{chay84, CFL, RE89, terman92}. 
In view of the complex bifurcation
structure of this class of problems, it is important to identify the universal
features pertinent to different dynamical patterns and transitions between them. 
In the present Letter, we describe statistical features of the irregular bursting
arising in a class of neuronal models close to the transition from
spiking to bursting. Prior to transition
to bursting, the systems in this class develop chaotic attractors, which
generate irregular spiking. The chaotic spiking gives rise to 
irregular bursting. The duration of bursts near the transition can be 
extremely long (see Figure \ref{f.1}). 
We analyze the statistics of the number of spikes and the interspike intervals within 
one burst as functions of the distance from criticality.

To describe our results, we use a three variable model of a 
bursting neuron introduced in \cite{IZH05}. The model is based on three 
nonlinear conductances: persistent sodium, $I_{NaP}$, the delayed rectifier, $I_K$, 
a slow  potassium $M$-current, $I_{KM}$, and a passive current, $I_L$.
In spite of a number of simplifications, 
the model captures essential features of a class
of more detailed physiological models and is representative 
for a class of square-wave bursters  \cite{IZH05}.
The latter are among most common models of bursting.
The following system of three differential equations describes the dynamics
of the membrane potential, $v$, and two gating variables $n$ and $w$:
\begin{eqnarray}\lbl{1}
C\dot v &=&F(v,n,w),\\\lbl{2}
\dot n &=& \frac{n_\infty(v)-n}{\tau_n},\\
\lbl{3}
\dot w &=& \frac{w_\infty(v)-w}{\tau_w},
\end{eqnarray}
where $F(v,n,w)=-g_{NaP}m_\infty(v)(v-E_{NaP})-g_Kn(v-E_K)-\gamma w(v-E_{K})-g_L(v-E_L)+I$;
$g_s$ and $E_s$, $\left(s \in \left\{NaP, K, L\right\}\right)$ are the maximal conductance
and the reversal potential of $I_s$, $s \in \left\{NaP, K, L\right\}$, respectively; and
$I$ is the applied current. The maximal conductance of $I_{KM}$, $\gamma$, 
is viewed as a control parameter.
The time constants $\tau_n$ and $\tau_w$ determine the rates of 
activation in the populations of $K$ and $KM$ channels. The 
steady state functions are defined by
$$
s_\infty (v) =\frac{1}{1+\exp\left(\frac{a_s-v}{b_s}\right)},\qquad s\in\left\{m,n,w\right\}. 
$$
The parameter values are given in the caption to Figure \ref{f.1}.

The analysis of the bursting neuron models like (\ref{1})-(\ref{3}) uses a
fast-slow decomposition \cite{rinzel87, IZH05, RE89}. Specifically, we note that the time constant
$\tau_w$ presents the slowest time scale in the dynamics of (\ref{1})-(\ref{3}).
Therefore, we view $\alpha=\tau_w^{-1}>0$ as a small parameter. In the limit as
$\alpha \rightarrow 0$, system (\ref{1})-(\ref{3}) is reduced to
a $2D$ fast subsystem (\ref{1}), (\ref{2}), where $w$ is viewed as a parameter.
\begin{figure}
\begin{center}
\epsfig{figure=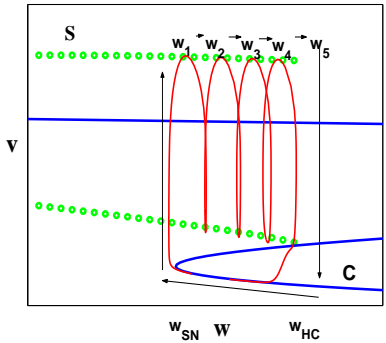, height=1.8in, width=2.0in, angle=0}
\end{center}
\caption{ The bifurcation diagram of the fast subsystem (\ref{1}) 
and (\ref{2}) with a superimposed
periodic trajectory induced by the slow equation (\ref{6}). 
Curves S and C indicate the families of periodic orbits and the fixed points respectively.
}
\label{f.2}
\end{figure}
For values of $w \in \left(w_{AH},w_{HC}\right)$ the fast subsystem has a family of 
stable periodic orbits, which is  born at the Andronov-Hopf  bifurcation at $w=w_{AH}$ and
terminates at the homoclinic bifurcation at $w=w_{HC}$.
A trajectory of the full system (\ref{1})-(\ref{3}) approaches the surface foliated by 
the periodic orbits of the fast subsystem, $S$ (Figure \ref{f.2}).
The evolution along $S$ is governed by the slow equation:
\begin{equation}\lbl{6}
\dot w=\alpha \left(w_\infty(v)-w\right),
\end{equation}
where the leading order approximation of $v(t)$ is obtained from the fast equations 
(\ref{1}) and (\ref{2}). 
When the trajectory hits the boundary of $S$ (i.e., when $w> w_{HC}$), 
it jumps down to the curve of stable fixed points $C$, the only attracting 
set of the fast subsystem 
for $w>w_{HC}$. Then it evolves along $C$ as 
shown in Figure \ref{f.2}
until it reaches the boundary of $C$ at $w=w_{SN}$, corresponding 
to a saddle-node (SN) bifurcation in the fast subsystem. This is 
followed by the jump back to $S$ and the oscillations in the fast
subsystem resume. This description accounts for one cycle of bursting
oscillation. 

\begin{figure}
\begin{center}
\epsfig{figure=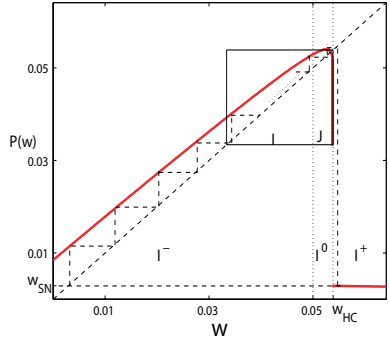, height=1.8in, width=2.0in, angle=0} 
\end{center}
\caption{The first return map for the slow variable $w$ for the value of control
parameter $\gamma=2.9375$ close to the critical value $\gamma^c$.}
\label{f.3}
\end{figure}

The description of bursting dynamics in a wide class of
neuronal models, including (\ref{1})-(\ref{3}), can be reduced to a single 
difference equation for the slow variable.
Indeed, since the state of the fast system is determined by the value of 
the slow variable $w$, it is sufficient to know how $w$ changes after each 
cycle of oscillations of the fast subsystem and after the period of quiescence:
\be\lbl{7}
w_{n+1}=P_\gamma\left(w_n\right),\; n=1,2,\dots.
\ee
This idea underlies the method of reduction of a class of models of bursting
neurons to one-dimensional maps proposed in \cite{M05}. 
Similar map-based approaches were used in \cite{rinzel_troy82a}
for studying bursting patterns in the Belousov-Zhabotinskii reaction and
in \cite{MC} for analyzing complex oscillatory regimes in a compartmental
model of the dopamine neuron.
In \cite{M05}, 
we provided the analytical description for $P_\gamma$, which was used to account 
for various spiking and bursting patterns and transitions between
in the class of models of excitable cells.
For the model at hand, the one-dimensional map is shown in Figure \ref{f.3}.
The bifurcation structure of the fast subsystem endows the map with
distinct structure: it is a piecewise continuous map with the boundary layer,
$I_0$, corresponding to the homoclinic bifurcation in the fast subsystem
(Figure \ref{f.3}).
There are two intervals of 
continuity in the domain of definition of $P_\gamma$, $I_1=I^-\bigcup I^0$ and 
$I_2=I^+$. The iterations of $P_\gamma$ over $I_1$ correspond to the changes of $w$
after each spike of voltage and the definition of
$P_\gamma$ over $I_2$ captures the mechanism of return to spiking after the
period of quiescence (Figure \ref{f.3}). In $I_2$, $P_\gamma\approx w_{SN}$ 
is almost constant 
and can be well approximated by a linear function with a small negative slope
(see Remark 5.3(b) in \cite{M05}).
By construction, $P_\gamma$ has a unique fixed point $\bar w_\gamma \in I_1$.
For small values of $\gamma$, $\bar w_\gamma$ is globally attracting.
The trajectory of (\ref{7}) stays in a small neighborhood of 
$\bar w_\gamma \in I_1$ after several iterations of $P_\gamma$. Therefore, the
value of
$w$ in (\ref{1}) and (\ref{2}) changes little and the fast subsystem
generates periodic spiking. The bifurcation scenario described in 
\cite{M05} 
for a model of a bursting neuron
applies to a wide class problems including (\ref{1})-(\ref{3}).
According to this scenario, for increasing values of $\gamma$
the fixed point $\bar w_\gamma$ loses stability through a period-doubling (PD)
bifurcation, giving rise to a stable period $2$ orbit. In the continuous
system (\ref{1})-(\ref{3}), this corresponds to the appearance of the stable 
periodic solution formed by clustered pairs of spikes, so-called doublets. 
Under further increase of $\gamma$, the numerical experiments show
other PD bifurcations of periodic orbits and, eventually, the 
dynamics of the discrete system (\ref{7}) becomes chaotic.
This scenario fits well with the numerical results reported for
related models \cite{chay84,CFL,M05}. 
While the maximum of $P_\gamma$ over $I_1$, $\bar P_\gamma=max_{w\in I_1}P_\gamma(w)$
remains less than $w_{HC}$, $P_\gamma(I_1) \subset I_1$, and the trajectories of (\ref{7})
are trapped in $I_1$. This means that, in this regime, the continuous
system exhibits (possibly chaotic) spiking. The transition to bursting takes
place at $\gamma=\gamma^c$:
\begin{equation}\lbl{8}
\bar P_{\gamma^c}=w_{HC}.
\end{equation}
For $\gamma >\gamma^c$, trajectories of (\ref{7}) may leave $I_1$.
However, for values of $\gamma$ just above $\gamma^c$, the window of escape,
$J_\gamma$, is very small (see Figure \ref{f.3}). Therefore, a trajectory of (\ref{7})
with large probability spends a long time in $I_1$ before escaping to $I_2$.
If, in addition, the dynamics of (\ref{7}) for $\gamma=\gamma_c$ has mixing
property, the transition to bursting  lies through regimes
of chaotic bursting with very long intervals of spiking appearing with large 
probability. Below we study the statistics of the number of
spikes and the interspike intervals within one burst for
(\ref{1})-(\ref{3}) near the transition to bursting.
\begin{figure}
\begin{center}
\epsfig{figure=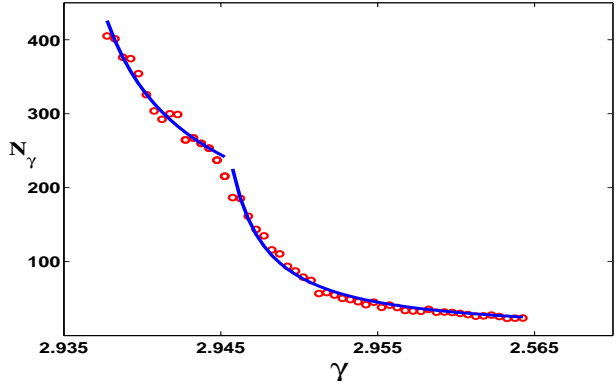, height=2.0in, width=3.2in, angle=0}
\end{center}
\caption{The mean values of the number of spikes, $N_\gamma$, are fitted
with $y=a\left(\gamma-\gamma^c\right)^{-1\over 2}, \left(\gamma<\bar\gamma\right)$ and 
$y=b\left(\gamma-\gamma^c\right)^{-1}, \left(\gamma>\bar\gamma\right)$, where 
$a\approx 25.42$, $b\approx 0.52$, $\gamma^c\approx 2.93419,$ and $\bar\gamma\approx 2.934343.$
}
\label{f.4}
\end{figure}

\begin{figure}
\begin{center}
\epsfig{figure=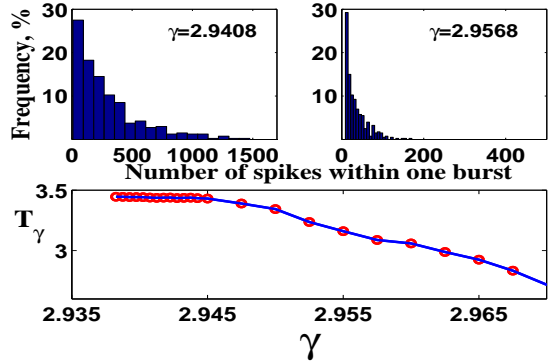, height=2.0in, width=3.2in, angle=0}
\end{center}
\caption{The histograms for the number of spikes within one burst 
and the mean values of the interspike intervals within one burst, $T_\gamma$. 
}
\label{f.5}
\end{figure}

Let $I=\left[w_0, w_{HC}\right],$ $w_0=\lim_{w\to w_{HC}-0} P_\gamma (w)$ (see Figure \ref{f.3})
and assume that at the critical value of the control parameter
$\gamma=\gamma^c$, map $P_{\gamma^c}: I \mapsto I$ has an invariant 
probability measure $\mu$ absolutely continuous to the Lebesgue measure on $I$.
In addition, we assume that $P_{\gamma^c}$ has the mixing property:
\begin{equation}\lbl{8a}
\lim_{n \to \infty} \mu\left(A \cap P_{\gamma^c}^{-n}(B)\right)=\mu(A)\mu(B),
\end{equation}
for all measurable sets $A,B\subset I$  \cite{sinai}.
Let $\gamma-\gamma^c>0$ be sufficiently small. 
Denote $\delta=\mu(J_\gamma),$ where $J_\gamma=\left\{w\in I:\; P_\gamma(w)>w_{HC}\right\}$
(Figure \ref{f.3}).
Then the expected value (with respect to the probability measure $\mu$) 
of the number of spikes within one burst is given by
\begin{equation}\lbl{9}
N_\gamma=\mu(J_\gamma)+\sum_{k=2}^{\infty}k\mu_k,
\end{equation}
where $\mu_k=\mu\left( P_\gamma^{-k+1}\left( J_\gamma\right) \bigcap 
\overline{ \bigcup_{i=0}^{k-2} P_\gamma^{-i}\left(J_\gamma\right)}\right)$ and $\overline{A}$ stands for
$I-A$.
Let  $m_k=\mu\left(\overline{\bigcup_{i=0}^{k-2}P^{-i}\left(J_\gamma\right)}\right)$ for $k\geq 2$.
Then by mixing property (\ref{8a}), for sufficiently large $k_0 \in \Na$,
\begin{equation}\lbl{11}
\mu_k\approx \delta m_k\approx \delta m_{k_0}\left(1-\delta\right)^{k-k_0},\; k>k_0.
\end{equation}
The combination of (\ref{9}) and (\ref{11}) yields 
\be\lbl{12}
N_\gamma=\Sigma_{k_0}+m_{k_0}\delta\sum_{k=1}^{\infty}\left(k+k_0\right)\left(1-\delta\right)^{k-k_0},
\ee
where $\Sigma_{k_0}$ stands for the first $k_0$ terms on the 
right-hand side of (\ref{9}). By taking into account,
$
\Sigma_{k_0}=O\left( 1 \right)
$
and
$\sum_{k=1}^{\infty}\left(k+k_0\right)\left(1-\delta\right)^k
=O\left(\delta^{-2}\right), 
$
from (\ref{12}) we obtain 
$
N_\gamma=O\left(\delta^{-1}\right).
$
In a small neighborhood around the point of maximum, the graph of $P_\gamma$ is 
to leading order quadratic. Therefore, the size of the window $J_\gamma$,
$\delta=\mu\left(J_\gamma\right)=O\left(\sqrt{\gamma-\gamma_c}\right),$
and $N_\gamma=O\left( \left(\gamma-\gamma^c \right)^{{-1\over 2}}\right)$.
To estimate $N_\gamma$ in a larger neighborhood of $\gamma^c$, we need to 
review certain facts about the structure of $P_\gamma$. 
It follows from the construction of $P_\gamma$
in \cite{M05}, that outside of the exponentially small neighborhood of $w_{HC}$,
$\tilde I$,
$P_\gamma$ can be approximated by a linear map.
This implies that for $\gamma>\bar\gamma=\gamma^c+O\left(e^{-C_1\over\alpha}\right)$
the size of $J_\gamma$ grows approximately linearly with $\gamma$, 
$\delta\approx O\left(\gamma-\bar\gamma\right)$ and 
$N_\gamma=O\left(\left(\gamma-\bar\gamma\right)^{-1}\right)$ for
$\gamma>\bar\gamma$.
Therefore, in the vicinity of $\gamma^c$,
there are two regions of qualitatively distinct asymptotic behaviors of $N_\gamma$
as a function of the distance from criticality. The numerical results shown in Figure
\ref{f.4} confirm this conclusion and clearly indicate the boundary
between these two regions, $\bar\gamma\approx 2.943$. 
The proposed mechanism for irregular bursting implies 
that near the transition to bursting, $N_\gamma$ has a geometric distribution, whose 
parameters are determined by the width of the window of escape, $J_\gamma$
(see Figure \ref{f.5}).

Next, we turn to estimating the expected value for the interspike intervals 
within one burst, $T_\gamma$.
For small $\gamma-\gamma^c>0$, we have
\begin{equation}\lbl{13}
T_\gamma \approx \int_{I-J_\gamma}T\left(w,\gamma^c\right)\,d\mu(w),
\end{equation}
where $T\left(w,\gamma\right)$ stands for the period of oscillations in the
fast subsystem (\ref{1}),(\ref{2}). By the Lebesgue-Besicovitch differentiation theorem \cite{EG}, 
$\int_{J_\gamma}T\left(w,\gamma^c\right)d\mu (w)=O\left(\mu\left(J_\gamma\right)\right)=O\left(\delta\right)$,
for small $\gamma-\gamma^c>0$. Since 
$\delta=O\left(\sqrt{\gamma-\gamma^c}\right)$ for $\gamma\in\left(\gamma^c,\bar\gamma\right)$,
in this region, we have $T_\gamma=T_{\gamma^c}-O\left(\sqrt{\gamma-\gamma^c}\right).$ 
To estimate $T_\gamma$ for $\gamma>\bar\gamma$, we again use the proximity of $P_\gamma$
to a linear map in $I-\tilde I$, which implies that 
$I-J_\gamma\approx \left(w_0,w_{HC}-O\left(\delta\right)\right)$
and $d\mu(w)\approx C_2dw$ in $I-\tilde I$ for some $C_2>0$. In addition, by the well known results 
of the bifurcation theory \cite{GH},
$$
T\left(w,\gamma^c\right)\approx T_0 -C_3\log\left(w_{HC}-w\right),\;
$$
where positive constants $T_0$ and $C_3$ are independent from $w$,
because the fast subsystem is close to a homoclinic bifurcation. Using these approximations
and (\ref{13}), we obtain
$$
T_\gamma\approx C_2 \int_{w_0}^{w_{HC}-\delta}T\left(w,\gamma^c\right)dw\approx
\bar T-C_4\delta\left|\ln\delta\right| +O(\delta),
$$ 
where positive constants $\bar T$ and $C_6$ are independent of $\delta$ and
$\delta=O\left(\gamma-\gamma^c\right).$ The numerical results in Figure \ref{f.5}
show that the asymptotic behaviors of $T_\gamma$ are different for $\gamma<\bar\gamma$
and $\gamma>\bar\gamma.$ The sublinear character of $T_\gamma$ in the latter region is
consistent with our estimate 
$T_\gamma-\bar T=O\left(\left(\gamma-\bar\gamma\right)
\left|\log\left(\gamma-\bar\gamma\right)\right|\right).
$
\begin{figure}
\begin{center}
\epsfig{figure=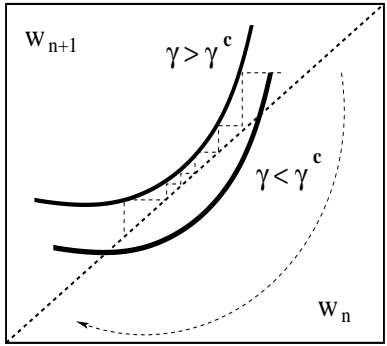, height=1.8in, width=2.0in, angle=0} 
\end{center}
\caption{
The mechanism of the transition to bursting through a SN bifurcation.
The SN bifurcation gives rise to a 
periodic orbit corresponding to a bursting solution. The global return mechanism
is schematically indicated by the dashed arc. 
}\label{f.6}
\end{figure}

The map-based approach to bursting employed in this paper reduces the problem of transition
from (tonic) spiking to bursting to the analysis of bifurcation scenarios in the families
of the first return maps. The first event in these scenarios is the loss of stability of the
fixed point corresponding to tonic spiking in the continuous system. For $1D$ maps,
 there are two co-dimension 1 bifurcations of fixed points: a SN and a PD bifurcation
\cite{GH}. The bifurcation scenario described in \cite{M05} and in the present paper  
originates from a PD bifurcation. A complimentary mechanism of the transition to bursting
is realized through a SN bifurcation (see Figure \ref{f.6}). 
It has been recently demonstrated for a model
for the leech heart interneuron \cite{SC04} (see also Remark 6.1(b) in \cite{M05}).
In analogy to the classification of excitability in spiking neuron models due to Rinzel and
Ermentrout \cite{RE89}, we propose to distinguish mechanisms of transition to bursting 
according to the underlying bifurcation mechanisms in the families of the first return maps:
{\bf Type I} transition is realized via a SN bifurcation of the fixed point (which coincides
with the global bifurcation of a periodic orbit, see Figure \ref{f.6});
{\bf Type II} transition is realized through the disappearance (deflation) of the chaotic
attractor (an attracting invariant interval with mixing property) preceded by a PD bifurcation. 
In both scenarios, the transition to bursting takes place after global bifurcations. However, the essential
part of the bifurcation mechanism, which determines the traits of the bifurcating bursting patterns,
in each case has a local character: a SN bifurcation in the Type I scenario and the deflation of the
chaotic attractor in Type II. The bursting patterns arising through these mechanisms 
possess well-defined statistical properties in terms of the mean burst duration and interspike 
interval distributions within one burst. In particular, the interspike interval distributions in Type II bursting 
patterns are characterized by high variability, whereas those in Type I are localized.
The mechanisms for transition to bursting in neuronal models are reminiscent to those
studied in the context of the transition to turbulence via intermittency \cite{LL}. In fact, the duration of
(regular) bursting near the transition in the {\bf Type I} scenario and that of the laminar
phases in the Type-I intermittency \cite{PM80} share the same asymptotics due to the proximity
to a SN bifurcation in both cases. In contrast, the statistical properties of the bursting patterns
in {\bf Type II} scenario are determined by the long intervals of irregular (chaotic) behavior and have not
been studied before.

The author thanks to anonymous referees for insisting on careful numerical verification of the 
analytical estimates presented in this paper. This work was partially supported by the National 
Science Foundation under Grant No. 0417624.

\end{document}